# Tuning Spin Current Injection at Ferromagnet/Non-Magnet Interfaces by Molecular Design

Angela Wittmann[1], Guillaume Schweicher[1], Katharina Broch[2], Jiri Novak[3], Vincent Lami[4,5], David Cornil[6], Erik R. McNellis[7], Olia Zadvorna[1], Deepak Venkateshvaran[1], Kazuo Takimiya[8], Yves H. Geerts[9], Jerome Cornil[6], Yana Vaynzof[4,5], Jairo Sinova[6], Shun Watanabe[10] and Henning Sirringhaus[1]

[1] *Optoelectronics Group, Cavendish Laboratory, University of Cambridge, JJ Thompson Avenue, Cambridge CB3 0HE, UK*

[2] *Institute for Applied Physics, University of Tübingen, Auf der Morgenstelle 10, 72076 Tübingen, Germany*

[3] *CEITEC MU and Faculty of Science, Masaryk University, 61137 Brno, Czech Republic*

[4] *Kirchhof Institute for Physics, Im Neuenheimer Feld 227, Heidelberg University, 69120 Heidelberg, Germany*

[5] *Centre for Advanced Materials, Im Neuenheimer Feld 225, Heidelberg University, 69120 Heidelberg, Germany*

[6] *Laboratory for Chemistry of Novel Materials, University of Mons, 7000 Mons, Belgium*

[7] *Institute of Physics, Johannes Gutenberg-Universit¨at, 55128 Mainz, Germany*

[8] *RIKEN Center for Emergent Matter Science, Wako, Saitama 351-0198, Japan*

[9] *Laboratoire de Chimie des Polymeres, Universite Libre de Bruxelles, 1050 Bruxelles, Belgium*

[10] *Department of Advanced Materials Science, School of Frontier Sciences, The University of Tokyo, 5-1-5 Kashiwanoha, Kashiwa, Chiba 277-8561, Japan*

There is a growing interest in utilizing the distinctive material properties of organic semiconductors for spintronic applications. Here, we explore injection of pure spin current from Permalloy into a small molecule system based on dinaphtho[2,3-b:2,3-f]thieno[3,2-b]thiophene (DNTT) at ferromagnetic resonance. The unique tunability of organic materials by molecular design allows us to study the impact of interfacial properties on the spin injection efficiency systematically. We show that both, spin injection efficiency at the interface as well as the spin diffusion length can be tuned sensitively by the interfacial molecular structure and side chain substitution of the molecule.

Spintronics is an important emerging communications technology, in which information processing is based not on the charge of an electron but on its inherent spin angular momentum. An indispensable building block for spin-based information processing are pure spin currents, a flow of electrons' spin an- gular momentum without a net flow of charge. Spin currents can be injected from ferromagnets into non- magnetic materials by the process of spin pumping, which involves excitation of magnetisation precession in the ferromagnet under ferromagnetic resonance (FMR) excitation conditions [1, 2]. There have been extensive theoretical and experimental investigations of spin pumping into inorganic materials [3–7] and several studies have emphasized the importance of structural properties and cleanliness of the interface for efficient spin injection [8–10]. However, most studies have been performed on non-epitaxial inter- faces prepared by sputter deposition and have been limited by the restricted possibilities of varying the interfacial properties in a controlled manner. Little is known, for example, how the atomic structure, orientation and packing at an interface influences key parameters for spin injection, such as the spin mixing conductance.

There has been recent emerging interest in organic semiconductors as spintronic materials, largely owing to the long spin relaxation times and long spin diffusion lengths [11, 12] that could be achieved in these materials due to their weak hyperfine and spin- orbit coupling interactions. This makes them conceivably suitable as non-magnetic spin conductors. However, organic materials also offer wide materials tunability and potentially the ability to study cleaner interfaces than it is possible in non-epitaxial inorganic materials because no covalent interfacial bonds need to be formed when a van-der-Waals bonded molecular semiconductor is deposited onto a ferromagnet. In this work, we report the first direct measurement of spin injection from a ferromagnetic metal into organic semiconductors (OSCs) by using the spin pumping technique and measuring the associated FMR linewidth broadening. We systematically compare spin injection into a series of small molecule OSCs with different conjugated cores and side chain substitutions and detect a very strong dependence of the spin mixing conductance on the molecular structure at the interface.

The magnetization dynamics of a ferromagnetic material (FM) under FMR condition are well described by the Landau-Lifshitz-Gilbert equation [13, 14]

$$\frac{d\boldsymbol{M}}{dt} = -\gamma \boldsymbol{M} \times \boldsymbol{H}_{eff} + \frac{\alpha}{M_s} \boldsymbol{M} \times \frac{d\boldsymbol{M}}{dt}, \qquad (1)$$

where $\gamma$ denotes the gyromagnetic ratio, and $M_s$ is the saturation magnetization. In addition to the precession of the magnetization **M** around the effective magnetic field **H**$_{eff}$, it also includes a damping term. The Gilbert damping parameter $\alpha$ is directly proportional to the width $\Delta H_{FM}$ of the microwave absorption of the FM when sweeping the magnetic field across FMR at a fixed microwave frequency. A FM can pump spins into an adjacent non-magnetic material (NM) at FMR. The effective damping is in- creased by $\Delta\alpha = \frac{\gamma}{\omega}(\Delta H_{FM/NM} - \Delta H_{FM})$ due to the loss of spin angular momentum from the FM into the NM resulting in an increase in linewidth $\Delta H_{FM/NM}$. A schematic illustration of spin pumping at FMR is shown in Fig. 1a. The spin mixing conductance $g_{eff}^{\uparrow\downarrow}$, a measure of the efficiency of injecting spins from an FM into an NM at FMR, can be experimentally determined by measuring the increase in linewidth of the absorption due to an additional NM layer using

$$g_{eff}^{\uparrow\downarrow} = \frac{4\pi\gamma M_s t_{NM}}{g\mu_B \omega}(\Delta H_{FM/NM} - \Delta H_{FM}), \qquad (2)$$

where $t_{NM}$ is the thickness of the NM, $\omega$ is the driving frequency, and $g$ and $\mu_B$ denote the electron $g$-factor and the Bohr magneton respectively [15]. When the thickness of the NM layer is smaller than or comparable with the spin diffusion length of the material, there is a backflow of spins due to a spin accumulation at the interface, opposing the injected pure spin current and diminishing the net spin injection. The dependence of the additional damping $\Delta\alpha$ on the thickness of the NM layer has been described as

$$\Delta\alpha = A\left(1 + \frac{B}{\tanh(t_{FM}/\lambda_S)}\right)^{-1}, \qquad (3)$$

where A and B denote material-dependent constants [3].

Here, we study spin injection from thermally evaporated Permalloy (Ni80Fe20, Py) into derivatives of dinaphtho[2,3-b:2,3-f]thieno[3,2-b]thiophene (DNTT) and benzothieno[3,2-b][1]benzothiophene (BTBT, see SI S9), two of the best performing small molecule OSCs so far [16, 17]. The inset in Fig. 1a shows a schematic diagram of the interface between the FM and DNTT. The chemical structure of the three derivatives of DNTT used in this study is shown in Fig. 1b. The core DNTT molecule has been modified by addition of two phenyl rings and C8 alkyl side-chains at the ends of the core for Ph- DNTT-Ph and C8-DNTT-C8 respectively.

In order to determine the linewidth broadening due to spin injection into the organic material, we measure the microwave absorption of each of the Py films (9 nm) before and after deposition of the small molecule in an electron spin resonance (ESR) cavity at 9.4 GHz while sweeping the in-plane magnetic field across FMR (see SI section S1). The measured derivative of the absorbed microwave power with respect to the applied magnetic field dI is compared in Fig. 2a for a bare Py film (black line) and the same sample after deposition of a 75 nm thick DNTT film on top (red). A small, but significant broadening is observed for the Py/DNTT bilayer with respect to the bare Py film. This broadening is shown more clearly in the insets, which display magnified plots in the vicinity of the maximum and minimum of the measured $\frac{dI}{dH}$ FMR curves. It can be seen that the peaks of the Py/DNTT bilayer shift to lower (higher) magnetic field below (above) the FMR demonstrating an increase in linewidth of the microwave ab- sorption. The increase in linewidth is illustrated more prominently in Fig. 2b, where the difference between the symmetric component of the absorption signal of the bare Py film and the same Py film after deposition of DNTT is shown. The broadening of the absorption in Py/DNTT has been measured reproducibly for nine samples, and the difference in linewidth is summarized in a histogram in Fig. 2c. In order to simplify direct comparison between different samples, $\delta\Delta H$ has been normalized to a linewidth of $\Delta H_{FM} = 21$ G. On average the linewidth increases by $\delta\Delta H = 0.54\pm0.05$ G.

Changes in interfacial structure due to slight oxidation of the Py surface during the deposition of the DNTT film, for instance, can cause distortion of the lineshape and linewidth of the microwave absorption, which is not related to spin injection [18, 19]. In order to exclude contributions to the linewidth broadening from this effect, we have performed this experiment with an additional gold layer between the FM and OSC. The thickness of the gold inter- layer was chosen such that it is thin enough not to affect spin injection but protect the Py film from oxidation [20]. We verified that the gold interlayer does not significantly change the linewidth broadening (see SI section S2). Therefore, we can conclude that this significant increase in linewidth is a direct manifestation of spin injection from Py into DNTT.

Making use of equation 2, we can estimate the spin mixing conductance to be $g_{eff}^{\uparrow\downarrow} = 3.35 \cdot 10^{18} m^{-2}$ for Py/DNTT. It is interesting to compare this value to that if Py/Pt interfaces, which is one of the most widely studied systems because Pt is an efficient spin sink layer. Our measured spin mixing conductance for Py/DNTT is only one order of magnitude lower than that of Py/Pt samples prepared under the same interface conditions (see SI section S3). Given the comparably small charge carrier concentration in the semiconducting DNTT and weak spin-orbit coupling, the spin injection efficiency is remarkably high. To the best of our knowledge, our experiment constitutes the first successful observation of linewidth broadening in an organic semiconductor. It may be considered as one of the cleanest observations of spin injection into an organic semiconductor, as the interpretation of such linewidth broadening is beautifully simple and unambiguous compared to other techniques, that rely on less direct methods for spin detection, such as magnetoresistance measurements. To investigate the influence of the molecular structure at the interface we have studied spin injection into all three derivatives of DNTT introduced above. Such a comparison is challenging because the interface properties can vary sensitively between different batches of Py. In order to ensure comparability, we measured linewidth broadening for the three mate- rials on the same batch of magnetic substrates. We used films with a thickness of 75 nm for all three materials assuming that this would be thick enough to neglect any effect of spin current backflow (see discussion of thickness dependence below). The change in linewidth (Fig. 3a) is comparable for DNTT and Ph-DNTT-Ph, where $\delta\Delta H = 0.48\pm0.09$ G, and considerably suppressed for C8-DNTT-C8 with $\delta\Delta H = 0.10\pm0.04$ G. The extracted linewidth broadening and spin-mixing conductance of the three derivatives of DNTT are summarized in Tab. I. There is clearly a strong dependence of the spin mixing conductance at the FM/OSC interface on the molecular structure at the interface.

Measuring linewidth broadening of the three derivatives of DNTT as a function of thickness ranging from 10 nm to 75 nm allows us to investigate potential effects of spin current backflow from the OSC into the FM. The absolute linewidths were found to vary from batch to batch and the number of samples needed to compare the thickness dependence of the three materials exceeded our experimental capacity of a single batch. Therefore, the thickness dependence was measured on one batch of Py for each material. To compare the different derivatives of DNTT, we then scaled the values of the thickness- dependent linewidth broadening to match the results obtained on the 75 nm thick films in Fig. 3a. The results are shown in Fig. 3b. From the gradual on- set of $\delta\Delta H$ for DNTT and Ph-DNTT-Ph with film thickness, we can estimate the spin diffusion length to be 40 nm for DNTT and 30 nm for Ph-DNTT-Ph. The signal for C8-DNTT-C8 is already saturated at 10 nm, implying that the spin diffusion length is considerably shorter. These results confirm our assumption above that in all three materials a thickness of 75 nm is sufficiently thick to saturate the linewidth broadening. Furthermore, these results show that the molecular structure at the interface has a pronounced effect not only on the spin mixing conductance, which is an interfacial property but also on the spin diffusion length, which is a bulk property.

As the spin mixing conductance is found to be highly dependent on the molecular structure, we have carefully characterized the structural properties of the organic films grown on Py substrates. For this, we have measured X-ray reflectivity (XRR) and grazing incidence X-ray diffraction (GID) of the three derivatives of DNTT to characterize the structure parallel and perpendicular to the substrate surface, respectively (see SI section S4). For the XRR-data, the positions of the observed out-of-plane Bragg- reflections agree well with the crystal structures of the small molecules reported in literature [21–23] for organic films with thickness ranging from 5 nm to 40 nm. In the GID measurements, we observe a slight expansion of the unit cell of 2% at most for all three molecules. Although this indicates the formation of a thin film phase, we emphasize that the growth mode of the three molecules does not differ from the bulk. The three DNTT derivatives adopt a 2D herringbone packing in-plane and stack edge-on with the long axis of the molecule approximately perpendicular to the substrate as illustrated for DNTT in Fig. 1a.

However, it is not sufficient to solely characterize the crystalline order and molecular orientation in the bulk of the films, but it is crucial to also check for the existence of any "wetting" layers at the FM/OSC interface, in which the materials could exhibit a different crystal packing (polymorphism) and preferential alignment of the unit cell versus the substrate [24, 25]. Therefore, we have investigated the topography of a 0.5 monolayer film of DNTT grown on a Py substrate using atomic force microscopy (AFM) (Fig. 4a). The profile is shown in Fig. 4b. The step height between the Py substrate and the areas covered by a monolayer of DNTT coincides with the length of a DNTT molecule (1.74 nm). Thus, we can exclude the formation of a layer with a substantially different molecular orientation, such as a face-on orientation at the interface and conclude that the DNTT molecules adopt a preferential alignment edge-on to the Py substrate in the bulk as well as at the interface.

In order to compare the electronic structure of the three derivatives of DNTT at the interface with Py, ultra-violet photoemission spectroscopy (UPS) depth profiling measurements were carried out on thin ( 40 nm) OSC films on Py following the methodology developed in [26] (see SI section S5). Fig. 4c shows the position of the highest occupied molecular orbital (HOMO) with respect to the Fermi level as a function of film thickness for the three compounds. The measurements show that no significant band bending can be observed for DNTT and Ph-DNTT-Ph throughout the layer. In the case of C8-DNTT-C8, we observe a variation in the energetic distance of the HOMO to the Fermi level, especially near the interface with Permalloy. The three molecules appear slightly p-type doped and the injection barrier at the interface with Py is on the order of 1.1 eV in all three materials with only small differences between the molecules which are close to the accuracy of the UPS technique (50 meV). The lack of band bending in DNTT and Ph-DNTT-Ph justifies our implicit assumption above of a uniform electronic structure and spin diffusion into the bulk of the films. In C8-DNTT-C8 one might argue based on the larger separation of the HOMO onset from the Fermi level in the bulk that the spin concentration in the bulk could be significantly lower than at the interface. This could influence the observed thick- ness scaling of the linewidth broadening. However, we cannot rule out that some variation in HOMO onset in C8-DNTT-C8 is an experimental artefact caused by, for example, charging of the film during the measurement.

To inject a spin current into a non-magnetic semi- conductor and carry it away from the interface there must be a sufficient concentration of spin carriers in the semiconductor. A rough estimate of the mini- mum carrier concentration in the semiconductor that is required is on the order of $10^{14}$-$10^{15}$ cm$^{-3}$ (see SI section S6). The origin of these spins is most likely due to either extrinsic doping or due to charge injection across the Schottky barrier at the OSC/Py interface. Taking into account the density of states broadening of the organic semiconductor, a carrier concentration of this magnitude for injection from the metal contacts is realistic even for a large injection barrier of 1.1 eV [27].

Previously, the FM/OSC interface has been observed to exhibit so-called spinterface effects due to interactions and hybridisation creating spin- dependent density of states at the interface [28, 29]. However, the experimental results presented here suggest that the spin mixing conductance depends mainly on the molecular structure rather than interfacial states in the material system we studied. In order to verify the role of the spinterface effects, we have simulated the electronic properties of the Py/DNTT interface for edge-on and face-on orientation of the molecule [30] and calculated the spin density in the cationic radicals [31, 32] (see SI section S8 and S7 respectively). As shown in Fig. 4d, the total atomic spin moment (ASM) is approximately 30 times smaller for edge-on oriented DNTT on Py (-0.009 $\mu_B$, black squares) compared to face-on orientation (-0.267 $\mu_B$, red circles). The intensity difference of the spin polarization of DNTT between the two different conformations can be observed clearly in the projected density of states (PDOS) at the interface (Fig. 4e and f). For edge-on orientation, the PDOS of DNTT is very similar to the one obtained for the isolated case and shows no difference between up and down spin. In contrast, there is a notable difference in up and down PDOS and a cancellation of the molecular gap in face-on DNTT. Hence, these simulations confirm that we would indeed not expect spinterface effects to play a dominant role in the spin injection experiments as DNTT adopts an edge-on rather than face-on orientation on Py.

Finally, in order to test the generality of the effect of alkyl side-chains on the spin injection efficiency, another similar small molecule system has been investigated. We have compared linewidth broadening for Py/ [1]Benzothieno[3,2-b]benzothiophene (BTBT) and Py/C8-BTBT-C8 bilayers. Analogously to DNTT, we can measure significant linewidth broadening in Py/BTBT and the spin injection efficiency is strongly suppressed due to the alkyl side chains in C8-BTBT-C8 (see SI section S9).

In conclusion, we have reported the first successful observation of FMR linewidth broadening due to spin pumping at an FM/small molecule OSC interface. The experimental technique of FMR linewidth broadening is a direct and simple experiment with unambiguous interpretation. Consequently, our observation may be regarded as one of the clearest demonstrations of spin injection into an organic semiconductor. The spin injection efficiency as quantified by the interfacial spin mixing conductance is found to be strongly influenced by the molecular structure at the interface. In molecules in which the spin density comes in close contact with the Py interface, we observe spin mixing conductance values that come to within an order of magnitude to that measured at interfaces of Py with inorganic metals, such as Pt. The technique also allows estimating the spin diffusion length in the direction perpendicular to the film plane. We obtain values of 30 40 nm in systems in which efficient spin injection is possible. Separation of the core of the molecule from the interface to the FM by a layer of alkyl side-chains sup- presses both the linewidth broadening as well as the spin diffusion length significantly. Our work demonstrates the importance of the structural properties at the interface with the FM to control and optimize the spin injection into the non-magnetic material. The vast possibilities of molecular design for organic small molecules combined with the tunability of the spin mixing conductance open the possibility of synthesizing tailor-made organic materials to optimize spin injection properties.

The authors would like to thank S. Schott for fruitful discussions. The small molecules Ph-DNTT-Ph and C8-DNTT-C8 were supplied by Nippon Kayaku. Funding from the Alexander von Humboldt Foundation, the European Research Council (ERC Synergy Grant SC2 No. 610115 and ERC grant agreement 714067, ENERGYMAPS) and the Grant Agency of the Czech Republic Grant No. 14-37427G is acknowledged. V. L. and Y. V. thank the Junior professor Program of the Baden-Wuerttemberg Ministry of Science, Research and Art for funding. G.S. acknowledges postdoctoral fellowship support from the Wiener-Anspach Foundation and The Leverhulme Trust (Early Career Fellowship supported by the Isaac Newton Trust).

# References

[1] Y. Tserkovnyak, A. Brataas, and G. E. W. Bauer, Physical Review Letters **88**, 117601 (2002).
[2] M. V. Costache, M. Sladkov, S. M. Watts, C. H. van der Wal, and B. J. van Wees, Physical Review Letters **97**, 216603 (2006).
[3] Y. Tserkovnyak, A. Brataas, and G. E. W. Bauer, Physical Review B **66**, 224403 (2002).
[4] R. Urban, G. Woltersdorf, and B. Heinrich, Physical Review Letters **87**, 217204 (2001).
[5] F. D. Czeschka, L. Dreher, M. S. Brandt, M. Weiler, M. Althammer, I.-M. Imort, G. Reiss, A. Thomas, W. Schoch, W. Limmer, H. Huebl, R. Gross, and S. T. B. Goennenwein, Physical Review Letters **107**, 046601 (2011).
[6] M. Tokac, S. A. Bunyaev, G. N. Kakazei, D. S. Schmool, D. Atkinson, and A. T. Hindmarch, Phys- ical Review Letters **115**, 056601 (2015).
[7] Z. Qiu, K. Ando, K. Uchida, Y. Kajiwara, R. Taka- hashi, H. Nakayama, T. An, Y. Fujikawa, and E. Saitoh, Applied Physics Letters **103**, 092404 (2013).
[8] M. B. Jungfleisch, V. Lauer, R. Neb, a. V. Chumak, and B. Hillebrands, Applied Physics Letters **103**, 2011 (2013), arXiv:arXiv:1302.6697v1.
[9] J. C. Rojas-Sánchez, N. Reyren, P. Laczkowski, W. Savero, J. P. Attane, C. Deranlot, M. Jamet, J. M. George, L. Vila, and H. Jaffrès, Physical Review Letters **112**, 1 (2014), arXiv:arXiv:1312.2717v1.
[10] J. M. Shaw, H. T. Nembach, and T. J. Silva, Phys- ical Review B **85**, 054412 (2012).
[11] S. Watanabe, K. Ando, K. Kang, S. Mooser, Y. Vaynzof, H. Kurebayashi, E. Saitoh, and H. Sirringhaus, Nature Physics **10**, 308 (2014).
[12] S.-J. Wang, D. Venkateshvaran, M. R. Mahani, U. Chopra, E. R. McNellis, R. Di Pietro, S. Schott, A. Wittmann, G. Schweicher, M. Cubukcu, K. Kang,Carey, T. J. Wagner, J. Siebrecht, D. Wong, I. Jacobs, R. Aboljadayel, A. Ionescu, S. Egorov,
Mueller, O. Zadvorna, P. Skalski, C. Jellett, M. Little, A. Marks, I. McCulloch, J. Wunder- lich, J. Sinova, and H. Sirringhaus, Under Review (2018).
[13] L. Landau, in Collected Papers of L.D. Landau (El- sevier, 1965) pp. 101–114.
[14] T. Gilbert, IEEE Transactions on Magnetics 40, 3443 (2004), arXiv:arXiv:1011.1669v3.
[15] O. Mosendz, J. E. Pearson, F. Y. Fradin, G. E. W. Bauer, S. D. Bader, and A. Hoffmann, Physical Re- view Letters 104, 046601 (2010).
[16] T. Uemura, K. Nakayama, Y. Hirose, J. Soeda, M. Uno, W. Li, M. Yamagishi, Y. Okada and J. Takeya, Current Applied Physics 12, S87 (2012).
[17] G. Schweicher, Y. Olivier, V. Lemaur, and Y. H. Geerts, Israel Journal of Chemistry 54, 595 (2014).
[18] M. R. Fitzsimmons, T. J. Silva, and T. M. Craw- ford, Physical Review B 73, 014420 (2006).
[19] N. Haag, S. Steil, N. Großmann, R. Fetzer, M. Cinchetti, and M. Aeschlimann, Applied Physics Letters 103, 251603 (2013).
[20] M. Isasa, E. Villamor, L. E. Hueso, M. Gradhand, and F. Casanova, Physical Review B 91, 024402 (2015).
[21] M. J. Kang, E. Miyazaki, I. Osaka, K. Takimiya, and A. Nakao, ACS Applied Materials & Interfaces 5, 2331 (2013).
[22] M. J. Kang, E. Miyazaki, I. Osaka, and K. Takimiya, Japanese Journal of Applied Physics 51, 11PD04 (2012).
[23] T. Yamamoto and K. Takimiya, Journal of the American Chemical Society 129, 2224 (2007).
[24] D. Kafer, L. Ruppel, and G. Witte, Physical Review B 75, 085309 (2007).

[25] A. O. F. Jones, B. Chattopadhyay, Y. H. Geerts, and R. Resel, Advanced Functional Materials 26, 2233 (2016).
[26] V. Lami, A. Weu, J. Zhang, Y. Chen, Z. Fei, M. Heeney, R. H. Friend, and Y. Vaynzof, Under Review (2019).
[27] N. I. Craciun, J. J. Brondijk, and P. W. M. Blom, Physical Review B 77, 035206 (2008).
[28] S. Sanvito, Nature Physics 6, 562 (2010).
[29] M. Cinchetti, V. A. Dediu, and L. E. Hueso, Nature Materials 16, 507 (2017).
[30] J. M. Soler, E. Artacho, J. D. Gale, A. Garcıa, J. Junquera, P. Ordejon and D. Sanchez-Portal, Journal of Physics: Condensed Matter 14, 2745 (2002).
[31] S. Schott, E. R. McNellis, C. B. Nielsen, H.-Y. Chen, S. Watanabe, H. Tanaka, I. McCulloch, K. Takimiya, J. Sinova, and H. Sirringhaus, Nature Communications 8, 15200 (2017).
[32] E. R. McNellis, S. Schott, H. Sirringhaus and J. Sinova, Physical Review Materials 2, 074405 (2018).

**Figures:**

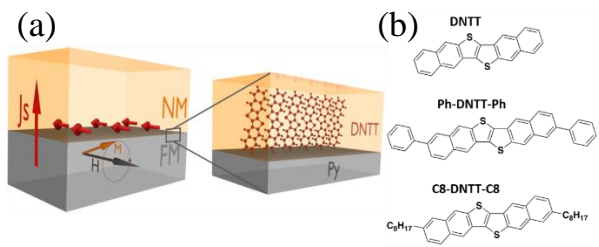

**Figure 1**: (a) Schematic illustration of injection of a pure spin current via spin pumping at FMR. The inset shows a zoomed-in view of the Py/DNTT interface. (b) Chemical structure of the three DNTT derivatives.

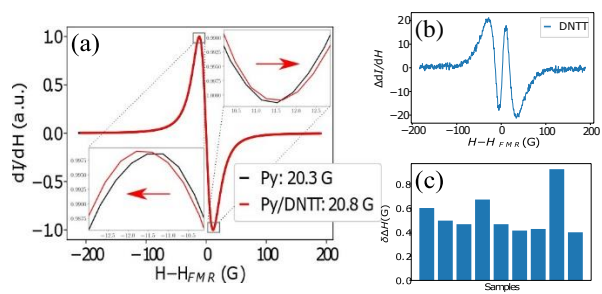

**Figure 2**: (a) Absorption spectra of the pristine Permalloy film (black) and after deposition of DNTT (red). The insets show the magnified region around the peaks clarifying the shift due to the increase in linewidth (red arrows). (b) Difference between the symmetric components of the normalized spectra. (c) Histogram of increase in linewidth for nine different samples.

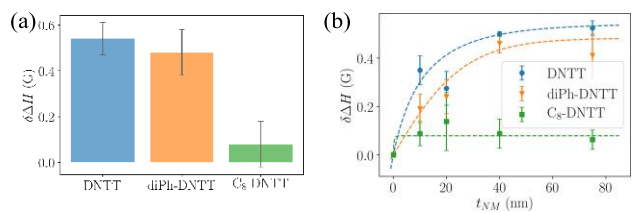

**Figure 3**: (a) Bar diagram illustrating the change in linewidth for Py/DNTT (blue), Py/Ph-DNTT-Ph (orange) and Py/C8-DNTT-C8 (green). (b) Linewidth broadening as a function of the thickness of DNTT (blue circles), Ph-DNTT-Ph (orange triangles) and C8-DNTT-C8 (green squares). From the gradual increase in signal with film thickness, the spin diffusion length can be estimated using the fit (dashed lines).

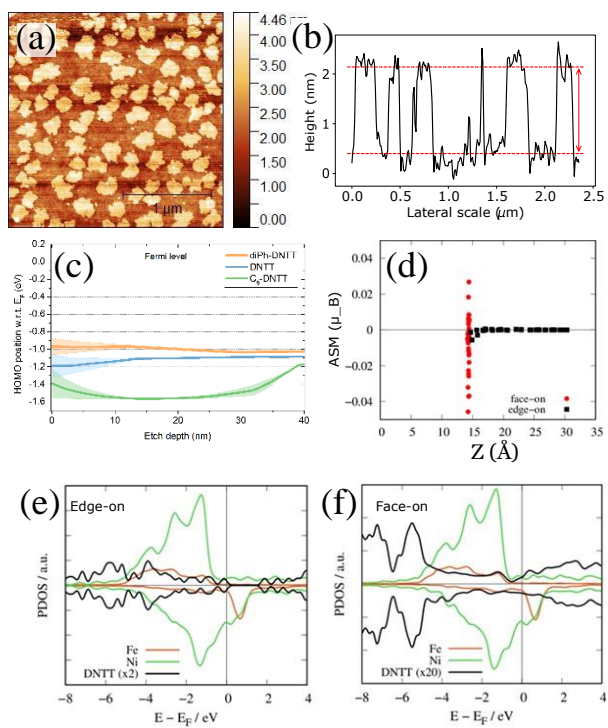

**Figure 4**: (a) AFM image of the topography of a 0.5 monolayer of DNTT on Py. The step height in the profile shown in (b) agrees with the length of the DNTT molecule. (c) HOMO position of DNTT (blue), Ph-DNTT-Ph (green) and C8-DNTT-C8 (red) with respect to the Fermi level. d) Spacial ASM distribution for edge-on and face-on oriented DNTT molecules on Py. (e) PDOS for edge-on and (f) face-on orientation of DNTT on Py.

**Table I**: Change in linewidth, spin-mixing conductance and estimate of spin diffusion length for the three derivatives of DNTT

|  | DNTT | diPh-DNTT | C8-DNTT |
|---|---|---|---|
| $\delta\Delta H$ (G) | $0.54 \pm 0.05$ | $0.48 \pm 0.09$ | $0.10 \pm 0.04$ |
| $g\uparrow\downarrow$ (m$^{-2}$) | $3.35 \cdot 10^{18}$ | $2.98 \cdot 10^{18}$ | $6.3 \cdot 10^{17}$ |
| $\lambda_s$ (nm) | 40 | 30 | 1 |